# Insight into the magnetism of a distorted Kagome lattice, $Dy_3Ru_4Al_{12}$, based on polycrystalline studies

Venkatesh Chandragiri, Kartik K Iyer and E.V. Sampathkumaran

*Tata Institute of Fundamental Research, Homi Bhabha Road, Colaba, Mumbai 400005, India*

Abstract

The layered compound with distorted Kagome nets, $Dy_3Ru_4Al_{12}$, was previously reported to undergo antiferromagnetic ordering below ($T_N=$) 7 K, based on investigations on single crystals. Here, we report the results of our investigation of *ac* and *dc* magnetic susceptibility ($\chi$), isothermal remnant magnetization ($M_{IRM}$), heat-capacity, magnetocaloric effect and magnetoresistance measurements on polycrystals. The present results reveal that there is an additional magnetic anomaly around 20 K, as though the Néel order is preceded by the formation of ferromagnetic clusters. We attribute this feature to geometric frustration of magnetism. In view of the existence of this phase, the interpretation of the linear-term in the heat-capacity in terms of spin-fluctuations from the Ru 4d band needs to be revisited. Additionally, in the vicinity of $T_N$, AC $\chi$ shows a prominent frequency dependence and, below $T_N$, $M_{IRM}$ exhibits a slow decay with time. This raises a question whether the antiferromagnetic structure in this compound is characterized by spin-glass-like dynamics. In contrast to what was reported earlier, there is a change in the sign of the magnetoresistance (MR) at the metamagnetic transition. A butter-fly-shaped (isothermal) MR loop (interestingly spanning over all the four quadrants) is observed at 2 K with distinct evidence for the magnetic phase co-existence phenomenon in zero field after travelling through metamagnetic transition field. The results on polycrystals thus provide additional information about the magnetism of this compound, revealing that the magnetism of this compound is more complex than what is believed, due to geometric frustration intrinsic to Kagome net.

## I. INTRODUCTION

Recently [1], single crystals of the compound, $Dy_3Ru_4Al_{12}$ [Ref. 2], crystallizing in $Gd_3Ru_4Al_{12}$-type hexagonal structure (space group: $P6_3/mmc$), were reported to exhibit interesting magnetic properties owing to certain structural features. The crystal structure [Refs. 2-3] contains planar layers of $Gd_3Al_4$ and puckered layers of $Ru_4Al_8$ stacked alternately along the $c$-axis. The magnetic atoms occupy the vertices of distorted Kagome nets and triangles, thereby favoring magnetic frustration for intersite antiferromagnetic coupling. Many interesting anomalies were reported [1], which require further understanding. These are: (i) *DC* magnetization ($M$) data taken with a magnetic field ($H$) of 10 kOe provided evidence for the onset of long range magnetic order at ($T_N=$) 7 K, Neutron diffraction results at 1.8 K reveal a complex antiferromagnetic structure. Interestingly, this magnetic transition was proposed to be first-order on the basis of heat-capacity ($C$) studies, in contrast to the expectation of second-order magnetic ordering. (ii) Magnetization jumps at different magnetic fields for different orientations of the single crystals favoring field-induced ferromagnetism have been found. There are corresponding jumps in magnetoresistance (*MR*). However, a *positive* magnetoresistance was observed even at high fields, which is unusual for the high-field ferromagnetic state. (iii) On the basis of large linear term in heat-capacity ($C$) above 7 K, it was proposed that Ru 4d band exhibits spin-fluctuations. We have carried out exhaustive investigations on polycrystalline samples by *AC* and *DC* magnetic susceptibility ($\chi$) with relatively low-fields, isothermal magnetization, heat-capacity and magnetoresistance studies. We obtain additional information, which are reported in this article. We think that the understanding of this compound with interesting crystallographic features (as mentioned above) favoring geometrically frustrated magnetism is still in its infancy.

## II. EXPERIMENTAL DETAILS

A polycrystalline specimen of the compound, $Dy_3Ru_4Al_{12}$, was prepared by melting together stoichiometric amounts of high purity (>99.9%) constituent elements in an arc furnace in an atmosphere of argon. The ingot was subsequently vacuum-annealed at 800 C for a week. The Rietveld-fitted powder x-ray diffraction pattern (Cu $K_\alpha$), shown in figure 1, confirms single phase nature of the sample. Scanning electron microscopy (SEM) was used to further characterize the sample. The temperature ($T$) dependence of *AC* as well as *DC* $\chi$ was measured (1.8 – 300 K) on a specimen of 8 mg with a commercial (Quantum Design) SQUID magnetometer. For *AC* measurements, the field employed was 1 Oe. $M(H)$ curves were also obtained at several temperatures. The electrical resistivity ($\rho$) as a function of $T$ in the presence of several magnetic fields and $\rho(H)$ curves at several temperatures were obtained with a commercial Physical Property Measurements System, PPMS (Quantum Design); heat-capacity was also studied with the same PPMS. Unless otherwise stated, all the measurements were performed for the zero-field-cooled (ZFC) condition of the specimen. An analogue, $Y_3Ru_4Al_{12}$, was also prepared by arc melting constituent elements to obtain lattice contribution to heat-capacity.

## III. RESULTS
### A. *Dc* magnetic susceptibility

Magnetic susceptibility measured in the presence of 5 kOe as a function of *T* in the form of $\chi$ and inverse $\chi$ is plotted in figure 2a. There is a distinct peak at 7 K, consistent with antiferromagnetic order [1]. From the Curie-Weiss fitting of the high temperature data, we find that the value of the paramagnetic Curie temperature ($\theta_p$) is $20\pm1$ K. The positive sign of $\theta_p$ implies the presence of ferromagnetic correlations, as emphasized in Ref. 1, though the compound orders antiferromagnetically. Above 7 K, inverse $\chi$ exhibits a deviation from the high temperature (>150 K) Curie-Weiss behavior as the *T* is lowered and is attributed to short-range magnetic correlations gradually building below 150 K. These are in good agreement with Ref. 1. However, measurements in small fields (20 and 100 Oe) reveal a bifurcation of the curves, obtained for ZFC and field-cooled (FC) conditions, below about 18 K (see Fig. 2b). We have measured χ in the presence of additional fields and the curves thus obtained tend to show dominant bifurcation around this temperature (Fig. 2b). The sharpness of the change in the slope of the curve around 20 K in figure 2b is large for low field measurements. [AC χ data (see below) also show a feature around 18 K]. These findings suggest that there is another magnetic feature above $T_N$. The effective moment obtained from the high temperature linear region (that is, ignoring temperature independent component) in the plot of inverse χ(*T*) is 10.5 ±0.03 $\mu_B$ per Dy, in close agreement with that expected for trivalent Dy (10.63 $\mu_B$). However, if we fit to modified Curie-Weiss form (i.e., adding a temperature independent component, $\chi_0$), one gets a value of 3.4 x $10^{-3}$ emu/mol for $\chi_0$ with a more satisfactory value for the effective moment.

## B. Isothermal *dc* magnetization

We have measured isothermal magnetization in 2K temperature steps between 2 – 90 K, in order to see the behavior of isothermal entropy change, *ΔS*= S(*H*)-S(0), at low temperatures, to offer support to the magnetic ordering behavior presented above. Derivation of *ΔS* employing Maxwell's thermodynamic relationship between magnetization and entropy and its relationship to the nature of magnetic ordering have been discussed at several places in the literature [4]. *M(H)* curves at selected temperatures are shown in figure 3. The values of ΔS obtained are shown in figure 4 for a change of the magnetic field from zero to a selected field (5, 10, 30, or 50 kOe). It is evident from this figure that the magnitude of ΔS gradually increases with decreasing *T*, attaining a peak, e.g., at a reasonably large value of 6.5J/kg-K for a final field of 50 kOe. It is notable that the peak appears at a slightly higher temperature compared to $T_N$ and the values remain in the positive quadrant in the plot of –ΔS versus *T* above $T_N$. The 'positive' peak in this plot with a rather magnitude of -*ΔS* for higher fields is typical of ferromagnetism [4]. We therefore conclude that there is a gradual formation of ferromagnetic clusters well above $T_N$ with decreasing temperature with its effect reflected more prominently near 20 K only in the low-field χ data. As the temperature is lowered below 10 K, following a sharp fall in the magnitude, there is a sign change and the observed negative sign is consistent with antiferromagnetic state. For low final fields of 5 and 10 kOe, there is a clear 'negative' peak in the plot at $T_N$, consistent with antiferromagnetism. Thus, this entropy data provides evidence for our conclusion in the previous paragraph. Finally, the *M(H)* curves below 7 K (see figure 3, for instance, for *T*= 2 K) show typical curvatures around 10 kOe, as though there is a magnetic-field-induced magnetic transition (towards ferromagnetic alignment); this is consistent with reports on single crystals [1]. For this reason, beyond this metamagnetic transition field, we do not observe a 'negative' peak in the plots in figure 4, for higher (final) fields.

Finally, considering that the above results offer evidence for the presence of both the antiferromagnetic and ferromagnetic interactions, we have looked for possible exchange-bias in the magnetization hysteresis loop. For this purpose, we cooled the sample from the paramagnetic state (150 K) to 1.8 K in the presence of 50 kOe and then took M(H) loop for a field variation sequence 50 kOe → -50 kOe → 50 kOe. We found that the hysteresis loop is symmetric with respect to origin, thereby ruling out any exchange bias. Therefore, we do not describe the material as the one in which antiferromagnetic clusters are embedded in ferromagnetic matrix or vice-versa.

## C. *AC* susceptibility

We have also carried out various studies to explore possible spin-glass anomalies [5]. We investigated the magnetic behavior by *AC* χ measurements with four frequencies (ν= 1.3, 13, 133 and 1333 Hz) (see figure 5). In the real part (χ′), apart from a peak at ~7 K, there is an additional prominent shoulder around 20 K. Though the frequency dependence of the peak around 7 K in χ′ is feeble as shown in the inset of figure 5, the ν-dependence is clearly visible in the imaginary part (χ″). Additionally, there is a distinct frequency dependence of the χ′ curves even below 7 K with the curves moving to a marginally higher temperatures with increasing frequency. This finding clearly implies glassy characteristics of the antiferromagnetic ordering. In order to explore other characteristics of spin-glasses [5, 6], we have measured *AC* χ in the presence of *DC* fields 1, 5 and 20 kOe, as it is expected that the *AC* χ signal should be suppressed [5] in the presence of a *DC* magnetic field (e.g., 1 kOe) for conventional spin glasses. In contrast to this expectation, the *AC* χ peaks around 7 K are not suppressed even in the presence of 5 kOe (see figure 5); higher fields are required to see the suppression, (which was verified by measuring with 20 kOe). This makes the magnetism of this compound different from conventional spin-glasses.

We observe a shoulder in χ′ around 18 K – below which ZFC-FC curves of low-field DC χ measurements separate out. Though the separation of the curves corresponding to different frequencies is visible, it is not sufficiently resolved to infer precise frequency dependence. In the same temperature region, χ″ shows a broad peak, though it is much weaker compared to the peak around 7 K. The temperature at which this weak peak in χ″ appears undergoes a shift with increasing ν, say by about 3 K for a variation of ν from 1.3 to 133 Hz. The shoulder in χ′ and the broad peak in χ″ vanish with the application of 1 kOe magnetic field. However, we do not attribute these features around 18 K to spin-glass freezing, as isothermal remnant magnetization ($M_{IRM}$) does not reveal any decay with time (*t*) at 12 K (*see below*). Possibly, the ferromagnetic clusters setting in near 18 K are characterized by a peculiar spin-dynamics, which is worth exploring further.

The right hand side inset in figure 5 shows the plot of ν versus inverse of $T_p$ (peak temperature in χ″ below 7 K). It is clear from this figure that Arrhenius relation (ν= $ν_0$exp(-$E_a$/$k_BT_p$), $ν_0$= pre-exponential factor, $E_a$= activation energy) is applicable. The value of $E_a$ is 50 K and the value of the pre-factor is of the order of 4 x $10^6$ Hz, which is about two orders of magnitude lower than that typical for spin-glasses [5]. We analyzed the AC χ results in terms of the conventional power law [5], associated with the critical slowdown of relaxation time, $τ/τ_0 = (T_f/T_g − 1)^{−zv}$. Here, τ = 1/2πν, $τ_0$ is the microscopic relaxation time, $T_g$ is the spin-glass transition temperature, $T_f$ corresponds to freezing temperature for a given *ν* and zν is the critical exponent. From this analysis, we obtain $T$g ≈ 3.6 K, zν ≈ 5.6 and $τ_0$ ≈ 5.6×$10^{−7}$ s, which corresponds to $ν_0$= 2 x $10^7$ Hz, closer to that expected for spin-glasses. The value of the critical exponent falls in the range expected for conventional spin glasses.

### D. Isothermal remnant magnetization and 'memory' studies

We have performed $M_{IRM}$ measurements at a few temperatures (1.8, 6 and 12 K) in the following manner: The specimen was cooled to a desired temperature in zero-field, a field of 5 kOe was applied, held for 5 minutes, and then the field was set back to 0 kOe. After this, at *t* = 0, the values of the remnant magnetization were found to be 0.16789 $μ_B$ / f. u, 0.05814 $μ_B$ / f.u. and 0.01943 $μ_B$ / f. u (where f.u. is formula unit) for 1.8, 6 and 12 K respectively. $M_{IRM}$ was then measured as a function time for about one hour. We find $M_{IRM}$ decays rather slowly with *t* at 1.8 and 6 K, and the curve can be fit to the stretched exponential form as shown in the figure 6, typical of spin-glasses. However, no such slow decay was found in the $M_{IRM}$ data above $T_N$ ; in fact, the value drops to much smaller values (as mentioned for 12 K above) remaining unchanged thereafter as soon as the magnetic-field is switched off, as shown for 12 K in the figure. This finding rules out onset of spin-freezing above $T_N$. We have also attempted to look for memory effect [6] in the magnetization data in the following manner. We cooled the sample to 4 K in zero field, waited for 4 h, then cooled to 1.8 K, applied a field of 20 Oe and measured magnetization in the range 1.8-50 K; we performed similar experiments without waiting at 4 K; we find that both the curves nearly overlap without any dip around 4 K in the difference curve. We found the same result when measured with 1 kOe as well. Thus, some characteristics of canonical spin-glasses could not be resolved in this material. Possibly the temperature window available (2 to 7 K) is too narrow to detect the same. Similar behavior was reported recently in some spin-chain systems [7].

### E. Heat-capacity

Temperature dependencies of heat-capacity measured in zero field, 5 and 20 kOe are shown in figure 7a. In zero field, there is a well-defined peak near 7 K, attributable to the onset of long-range magnetic ordering, as reported in Ref. 1. The fact that the peak temperature is suppressed in a small magnetic field (5 kOe) is consistent with antiferromagnetic ordering. However, the $T^3$-dependence of $C$ expected for antiferromagnets well below $T_N$ is not observed; instead, below 4 K, $C$ is dominated by a term linear in $T$ (also for 5 and 20 kOe, see figure 7). It is possible that such deviations are due to magnetic Brillioun-zone boundary gaps and/or interference from competing ferromagnetic interaction. Such behavior is not uncommon, for instance, as observed in $EuCu_2As_2$ [8] in which case there is a delicate balance between ferromagnetism and antiferromagnetism. This peak undergoes broadening for a magnetic-field of 20 kOe. But the peak for this field shifts marginally to a higher temperature by about a degree. We believe that this is due to dominating contributions of ferromagnetic clusters (setting in below 18 K), as evidenced by –ΔS data presented above. The fact that there is no heat-capacity anomaly around 18 K, even in the presence of external fields (measured till 100 K), establishes that there is no well-defined long-range magnetically ordered state above $T_N$. The plot of $C/T$ versus $T^2$ (see inset of figure 7b) is linear in the range 10-20 K. The linear term (γ) derived from this plot is rather large as in Ref. 1, but the magnitude observed here is nearly twice (~950 mJ/mol $K^2$ in contrast to 500 mJ/mol $K^2$ reported in Ref. 1) in zero field. Possibly magnetic cluster formation near 18 K is more facilitated in polycrystalline specimens due to crystallographic defects so as to result in more enhanced linear term. In the presence of 5 (20) kOe, this value undergoes a marginal change only to 1000 (1050) mJ/mol $K^2$. The heavy-fermion-like behavior was earlier attributed to spin-fluctuations from Ru 4d band in Ref. 1. We believe that this interpretation of the apparent linear term needs to be revised in light of the existence of the 18 K magnetic phase, revealed in this study. In this connection, it may be recalled that any magnetic precursor effect has been shown to mimic heavy-fermion behavior in Gd alloys [9]. From the linear region in the inset of figure 7, we attempted to obtain a Debye temperature ($θ_D$) and the value is found to be ~255 K, with negligible dependence on $H$. This value is close to that obtained from electrical resistivity data on single crystals in Ref. 1. There is a very weak deviation of the curve for 20 kOe from the zero-field curve in the paramagnetic state (see figure 7a). This deviation was found to gradually diminish with increasing temperature, merging with the zero-field curve around 50 K (not shown).

In order to understand magnetic entropy ($S_m$) at $T_N$, it is necessary to obtain the lattice contribution to heat-capacity. For this purpose, we measured $C(T)$ for the Y analogue, $Y_3Ru_4Al_{12}$ and the curve is included in figure 7a. The value of γ for the Y compound obtained from the linear region in the plot of $C/T$ versus $T^2$ (Fig. 2b inset) is found to be ~38 mJ/mol $K^2$ and the value of $θ_D$ is found to be ~300 K. We have followed the procedure suggested by Blanco et al [10] to account for the mass differences between Y and Dy compounds, while deriving phonon part for the Dy compound. The magnetic contribution ($C_m$) to heat capacity obtained by subtracting the phonon part is plotted in figure 7b in the form of $C_m/T$ versus $T^2$ for $H$= 0, 5 and 20 kOe at low temperatures and a dashed line is drawn through the linear region to infer γ values. The zero-field and in-field $S_m$ curves derived from $C_m$ are plotted in figure 7c. The value of $S_m$ associated with the magnetic transition at $T_N$ is found to be about 12 J/mol K (in the absence of external field), which is far less than that expected for crystal-field split doublet ground state value (17.3 J/mol K). This finding, agreeing with Ref. 1, further suggests the presence of additional magnetic features above 7 K. With the application of magnetic-field, $S_m$ at $T_N$ marginally changes. However, the dependence on $H$ appears to be weakly non-monotonic at $T_N$. This tendency persists with the increase of temperature. Possibly, an interplay among Zeeman effect, formation of ferromagnetic clusters and crystal-field effects could be responsible for such a field dependence of $S_m$.

Finally, we obtained isothermal entropy change for a change of $H$ from zero field to 20 kOe, and the plot as a function of $T$ obtained was found to be in agreement with that from magnetization data discussed above.

### F. Magnetoresistance

Single crystals of this compound revealed [1] an interesting magnetoresistance anomaly. MR, defined as [$\rho(H)-\rho(0)$]/$\rho(0)$], was shown to exhibit a jump with a positive sign at the metamagnetic transition field without any change in sign even at high fields, which was attributed to anomalous conduction electron contributions. Though such field-induced transitions should ideally result in negative magnetoresistance, such a positive MR behavior has been reported for some systems. Such a feature was attributed to 'inverse metamagnetism' [11]. We have therefore investigated how our sample behaves in this respect. $\rho(T)$ plots obtained in the absence and in the presence of magnetic fields ($H$=0, 30, 50 and 100 kOe) are shown in figure 8a. The absolute values of residual resistivity and the residual resistivity ratio, $\rho(300K)/\rho(0)$, are comparable to those in single crystals (e.g., along [100] direction), thereby attesting to the quality of our polycrystals. The shape of the curve (for $H$= 0) in the entire temperature range of investigation (including the drop at $T_N$) also compares quite well with that of single crystals. We show the data in the inset in the vicinity of magnetic ordering temperature in an expanded form to make the features more transparent. The point to be emphasized in figure 8 is that the zero-field and in-field curves separate well above $T_N$ (below about 100 K), similar to the deviation from Curie-Weiss region of $\chi$. The sign of the MR above ~7 K is negative and the magnitude increases with decreasing temperature (see Fig. 8b), similar to that in many rare-earth alloys in which magnetic precursor effects dominate before long-range magnetic order sets in [12]. Notably, the MR peaks above 10 K just as –$\Delta S$, and not at $T_N$; in addition, there is a sign reversal for $H$= 30 and 50 kOe at $T_N$, and the positive sign below $T_N$ is not inconsistent with antiferromagnetism. For higher fields, e.g., $H$= 100 kOe, the sign of MR remains negative, in contrast to Ref. 1. In order to get a better picture, we have measured MR as a function of $H$ at selected temperatures (2, 5, 8, 15 and 35 K; see figure 9). The shape of the MR curve is qualitatively different from that reported for single crystals [1] at 2 and 5 K. That is, in the virgin curve of MR($H$), there is an upturn at the metamagnetic transition field around 10 kOe and then a peak appears; the value subsequently undergoes a steep fall, exhibits a shoulder around 40 kOe and then changes sign (to negative) just beyond 50 kOe. This observation, in sharp contrast to that seen in single crystals, is in better agreement with the generally expected behavior that, after crossing the metamagnetic transition field, the ferromagnetically aligned spins should favor a negative MR. The reason for the different behavior from the single crystal is not clear at present. To clarify this, it may be worthwhile to measure the MR for additional orientations to look for negative MR values at high fields. Alternatively, it is not clear whether there is an unusual influence of grain boundaries (which are expected to be larger in numbers in polycrystals) on the MR. However, the quality of our polycrystals is comparable to that of the single crystal, as evidenced by residual electrical resistivity. On reversing the field to zero, the MR is found to be hysteretic, with the zero-field value after travelling through the metamagnetic transition field higher than for the virgin state. Thus, the virgin curve lies outside the envelope curve and this is typical of first-order (field-induced) phase transitions. This implies that the non-virgin zero-field state is characterized by a magnetic phase co-existence with a mixture of virgin and high-field states. A butterfly-shaped loop [13], *spanning over all four quadrants* in the MR($H$) plot is shown in figure 9a. The MR($H$) loop at 5 K is similar to that at 2 K, except that the virgin state resistivity is restored following field cycling. In the high-field range, 50-80 kOe, the plots show a weak hysteresis, which could be due to the existence of another metamagnetic transition (as shown in *M(H)* data for the geometry *H//[120]* in Ref. 1). On crossing $T_N$ (see inset of figure 9b for the curves at 8, 15 and 35 K), the sign of the MR remains negative, and the magnitude of the MR tends towards large values with increasing *H*. The reason for the flat region around 40-50 kOe in the 8 K plot is possibly an artifact of proximity to $T_N$.

## IV. Discussion

The magnetic nature above and below $T_N$ requires a special attention and therefore this section is devoted to a discussion summarizing essential findings from various measurements.

One of the major outcomes of this work based on low-field $\chi$ (both in AC and DC) is that, in addition to the long-range magnetic ordering at 7 K, there is an additional magnetic anomaly around 20 K. Since value of $\theta_p$ is closer to the temperature where this new feature appears and the sign of $\theta_p$ is positive, ferromagnetic correlations are likely responsible. Since *C(T)* is featureless above 7 K, this anomaly can not be attributed to long range magnetic ordering. Therefore, we conclude that ferromagnetic clusters form below 20 K, well before long range antiferromagnetic ordering sets in. The sign of the ΔS data presented in Section III.B above and below $T_N$ is consistent with this picture. The appearance of such a collective phase before long range ordering could be a manifestation of geometrical frustration, as proposed in Ref. 14. The AC χ features are rather weak around this characteristic temperature, but the ν-dependence is clearly discernable; since $M_{IRM}$ decay is not observed, we conclude that this ν-dependence does not arise from spin-glass dynamics, but may be due to domain dynamics of ferromagnetic clusters, as reported for $Sr_3CuRhO_6$ [15]. Such a magnetic precursor effect clearly manifests in a large linear term in heat-capacity.

There is a distinct evidence for the spin-glass-like spin-dynamics as inferred from the frequency dependence of AC χ as well as in $M_{IRM}$ at the onset of long-range magnetic order below 7 K. Any interpretation based on conventional spin-glass freezing is in sharp contrast to antiferromagnetism inferred from the neutron diffraction pattern at 1.8 K in Ref. 1 as well as a strong peak near 7 K in $C(T)$ and positive MR at low-fields reported here. The absence of exchange-bias (as measured at 1.8 K) excludes an explanation in terms of a ferromagnetic region (e.g., from surface) embedded in an antiferromagnetic matrix (or vice-versa) at such low temperatures. There are different ways of reconciling these experimental facts: Since the specimen studied here is polycrystalline, there is a possibility that the uncompensated spins at the surface of the grains lead to weak ferromagnetism (or even a spin-glass). While it is difficult to rule out this surface effect, it is also not easy to estimate this contribution, given irregularities in grain shape and size. However, for comparison, we have also carried out [16] studies on the Gd analogue and we did not find spin-glass-like features. The SEM images of the arc-melted ingots of both the compounds suggest similar-sized (<30 nm) grains. Therefore, we are tempted to conclude that the spin-glass-like anomalies observed for Dy case alone could be intrinsic to the grains, as a manifestation of geometrical frustration. One can propose that the antiferromagnetism observed by neutron diffraction is of a cluster-type and therefore one can invoke the idea that the intercluster interaction strength is random to influence AC and DC χ. A similar geometrically frustrated antiferromagnetic system, $Ca_3Co_2O_6$, undergoes complex changes in magnetic structure in an unprecedentedly large time scale of several hours [17] with a significance frequency dependence of AC χ. It would therefore be interesting to subject the present compound to detailed neutron diffraction studies as a function of temperature and time to probe intra-grain spin-glass-like slow spin-dynamics, and also to explore whether the diffraction patterns contain a diffused component arising from small ferromagnetic clusters.

Finally, the difference between the Gd and Dy compounds with respect to spin-glass-like features in the same family is surprising. It is not clear whether the anisotropy of the orbitals involved in magnetism (in the case of Dy) plays a subtle role to determine the existence of such magnetic features.

## V. **Summary**

An investigation of $Dy_3Ru_4Al_{12}$ in its polycrystalline form reveal new anomalies:

(i) There is a broad magnetic feature setting in around 20 K, preceding the first-order magnetic transition at 7 K with lowering temperature, in the low field susceptibility data. The appearance of such a collective phase before long range order supports the proposal of Jaubert et al [14] for one of the manifestations of geometrical frustration.

(ii) The existence of the magnetic feature between 7 and 20 K may be responsible for the large linear term in heat-capacity, which was earlier attributed to spin fluctuations from Ru 4d band.

(iii) Well below $T_N$, the magnetoresistance undergoes the expected change in sign from positive to negative (with increasing magnetic-field) after the initial metamagnetic transition, unlike that seen in single crystals [1], and an anomalous butterfly-shaped magnetoresistance loop spanning all the four quadrants in the *MR(H)* plot could be seen. Signatures of magnetic phase coexistence are observed in the *MR(H)* plot at 2 K.

(iv) Characteristics of spin-glass freezing (frequency dependent real and imaginary parts of *ac* $\chi$, bifurcation of low-field *dc* $\chi$ curves and decay of $M_{IRM}$) were found in the magnetically ordered state. These features, possibly a manifestation of geometrically frustrated magnetism, are interpreted in terms of spin dynamics within the antiferromagnetic phase in this system.

In short, this compound appears to be a geometrically frustrated magnetism with rich physics, worth pursuing studies with other experimental methods. .

Figure 1:
X-ray diffraction (Cu K$_\alpha$) pattern for Dy$_3$Ru$_4$Al$_{12}$. Rietveld fitting, along with fitted parameters and lattice constants, is also included.

Figure 2:

(a) Magnetic susceptibility and inverse susceptibility as a function of temperature measured in a field of 5 kOe for Dy$_3$Ru$_4$Al$_{12}$. The continuous line in the inverse $\chi$ plot represents modified Curie-Weiss fit above 150 K. (b) Inverse susceptibility as a function of temperature below 50 K measured in various fields; in the inset, zero-field-cooled and field-cooled susceptibility taken in a low-field is plotted to show the bifurcation of the curves.

Figure 3:
Isothermal magnetization curves (forward cycle only) at selected temperatures (2, 5, 10, 14, 20, 30, 40, and 50 K) for Dy$_3$Ru$_4$Al$_{12}$.

Figure 4:
Isothermal entropy change, ($-\Delta S = S(0)-S(H)$) ($H= 5, 10, 20, 30,$ and 50 kOe) for selected final fields for Dy$_3$Ru$_4$Al$_{12}$. The lines are drawn through the data points.

Figure 5:
Real and imaginary parts of *ac* susceptibility measured with various frequencies ($\nu= 1.3, 13, 133$ and 1333 Hz) for Dy$_3$Ru$_4$Al$_{12}$ in the absence and in the presence of external *dc* magnetic fields (0, 1 and 5 kOe). For the sake of clarity, the curves for 1.3 and 1333 Hz only are shown for the in-field data. In left data, the plots around the peaks are shown in an expanded form. The right inset shows Arrhenius plot.

Figure 6:
Isothermal remnant magnetization behavior as a function of time for Dy$_3$Ru$_4$Al$_{12}$ at three temperatures. The curves for 1.8 and 6 K could be fitted to an exponential form shown at the curves.

Figure 7:
(a) Temperature dependence of heat-capacity in the absence and in the presence of magnetic fields (5 and 20 kOe) for Dy$_3$Ru$_4$Al$_{12}$: The curves are obtained by drawing a line through the data points. The $C(T)$ curve for the Y analogue and lattice part for the Dy compound derived from this curve are also shown.. (b) Magnetic contribution ($C_m$) to heat-capacity divided by $T$ versus $T^2$ for zero field for the Dy compound in the low temperature range and a dashed line is drawn through the linear region. Inset shows $C/T$ as a function $T^2$ for the Y compound in the low temperature linear region. (c) Magnetic entropy as a function of temperature for $H= 0, 5$ and 20 kOe below 20 K.

Figure 8:
(a). Electrical resistivity as a function of temperature for Dy$_3$Ru$_4$Al$_{12}$ in the absence and in the presence of magnetic fields (30, 50 and 100 kOe). Inset shows the low temperature data in an expanded form. (b) Magnetoresistance as a function of temperature derived are plotted.

Figure 9:
Magnetoresistance as a function of magnetic field (in the field range -140 to 140 kOe) for Dy$_3$Ru$_4$Al$_{12}$ at (a) 2 K with the inset showing low-field region in an expanded form, and (b) 5 K. In the inset of (b) the behavior at 8, 15, and 35 K is shown in the positive quadrant only. The numericals and arrows are placed on the curves to show the direction of change of magnetic field.

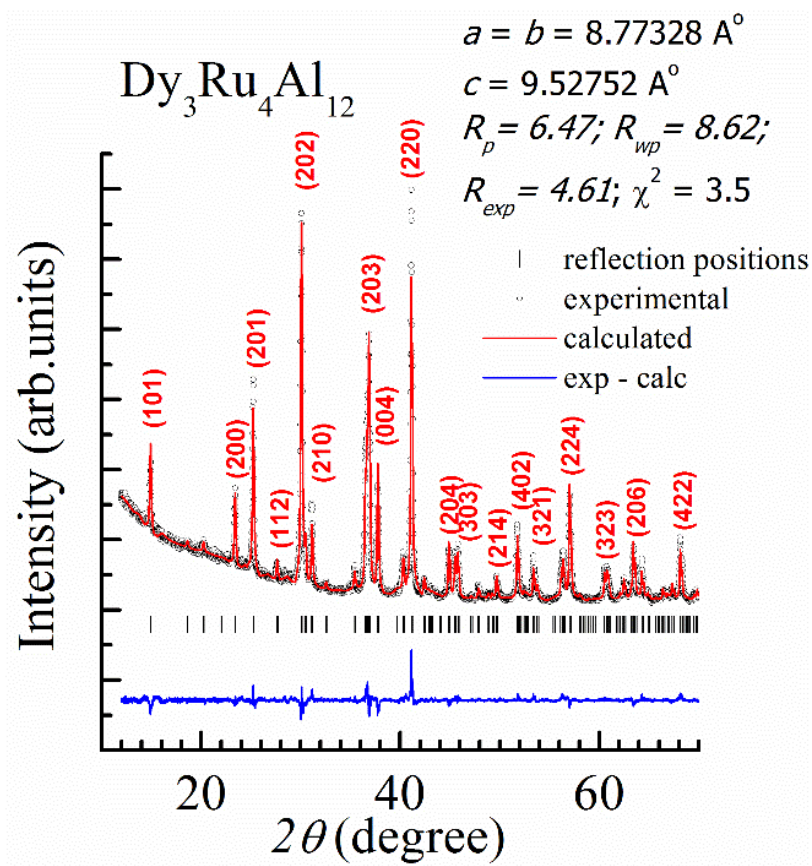

FIGURE 1

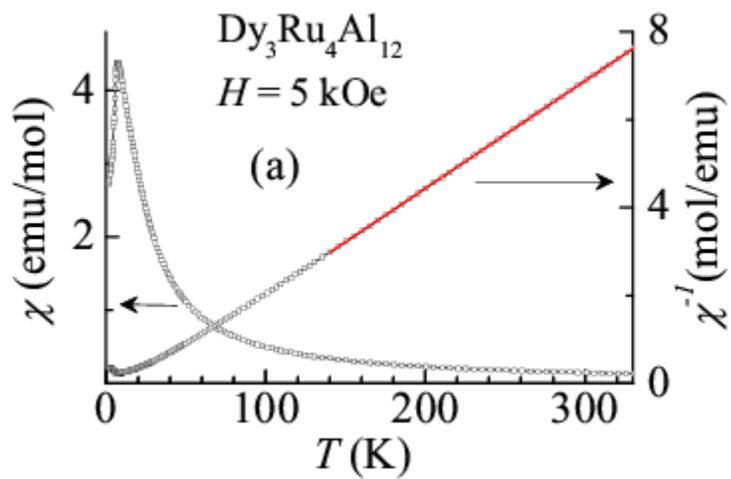

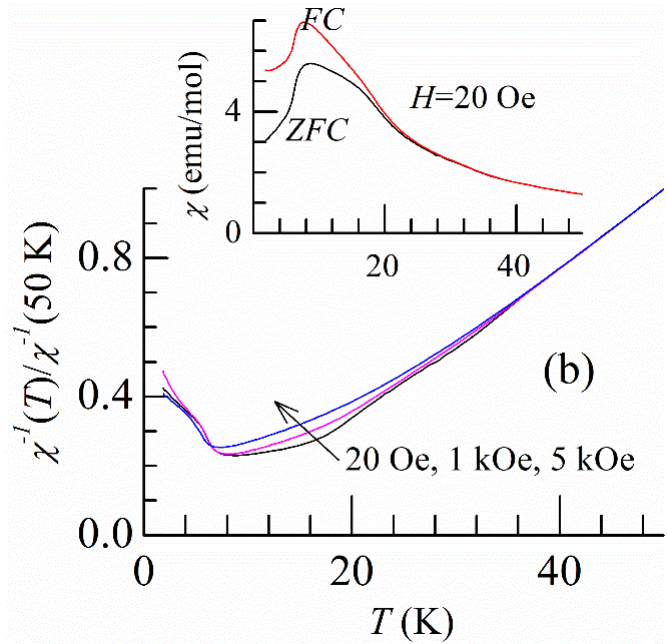

FIGURE 2

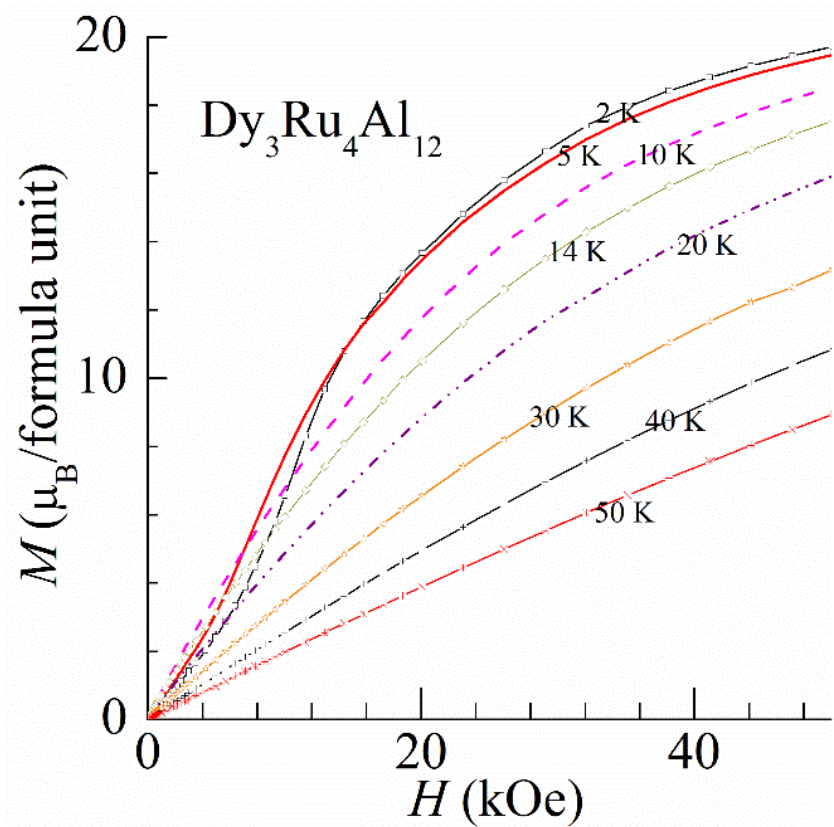

FIGURE 3

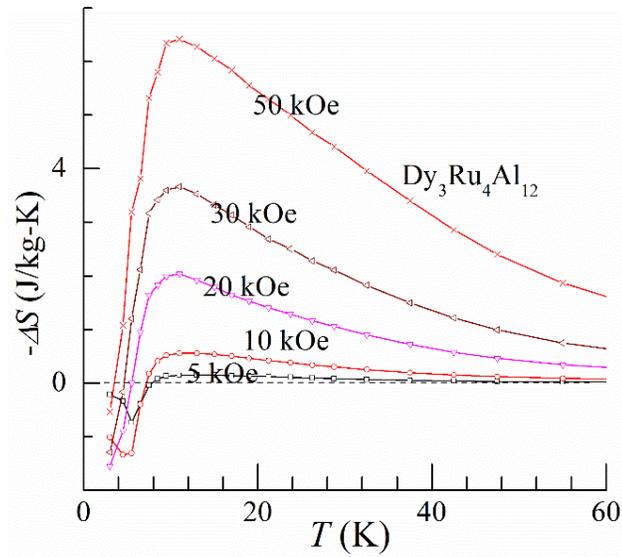

FIGURE 4

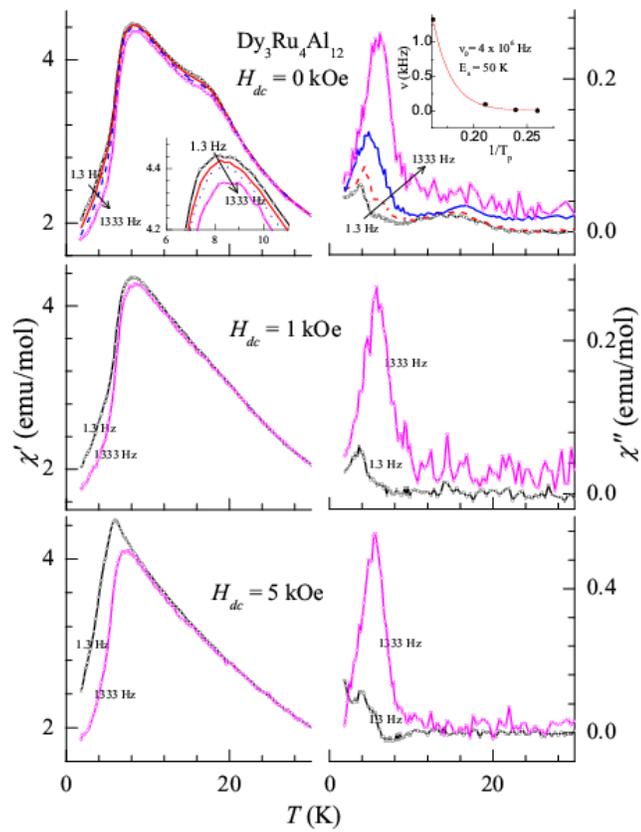

FIGURE 5

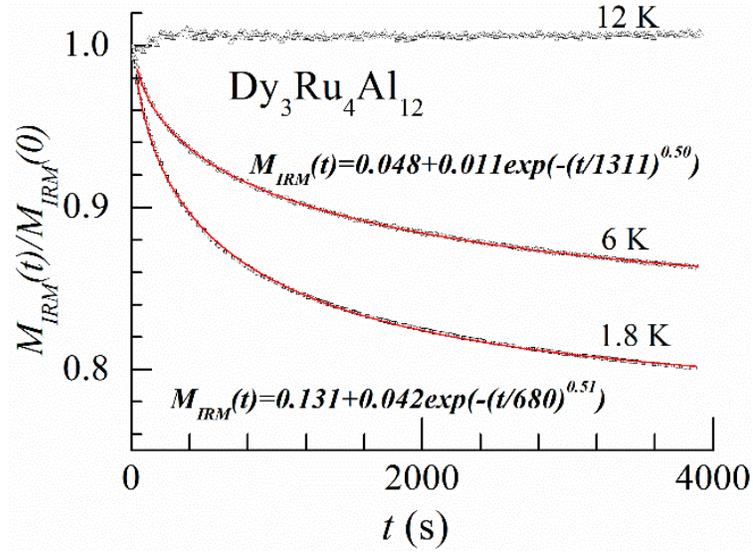

FIGURE 6

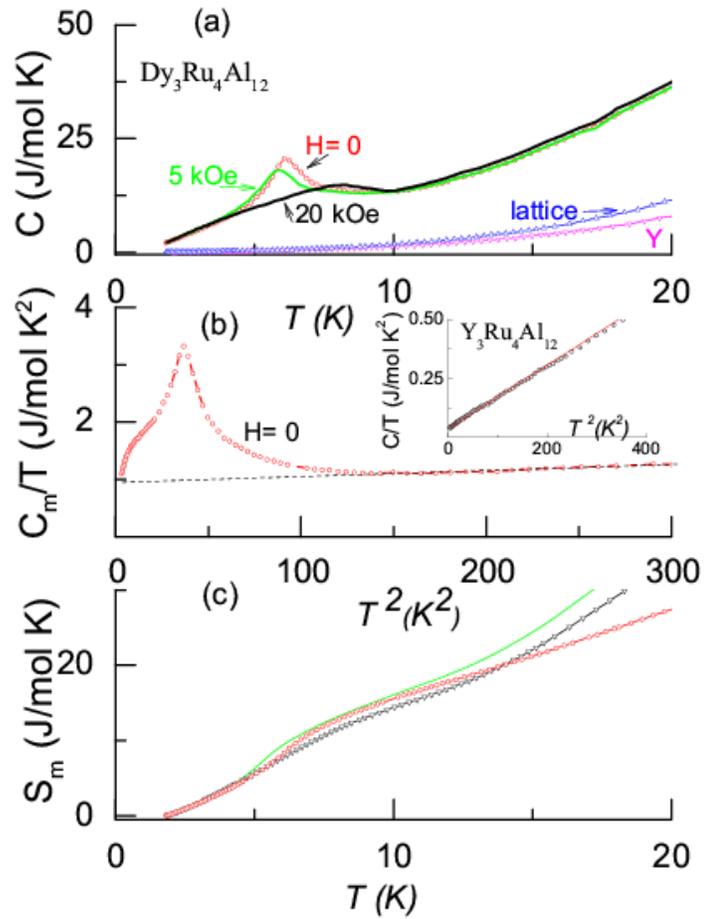

FIGURE 7

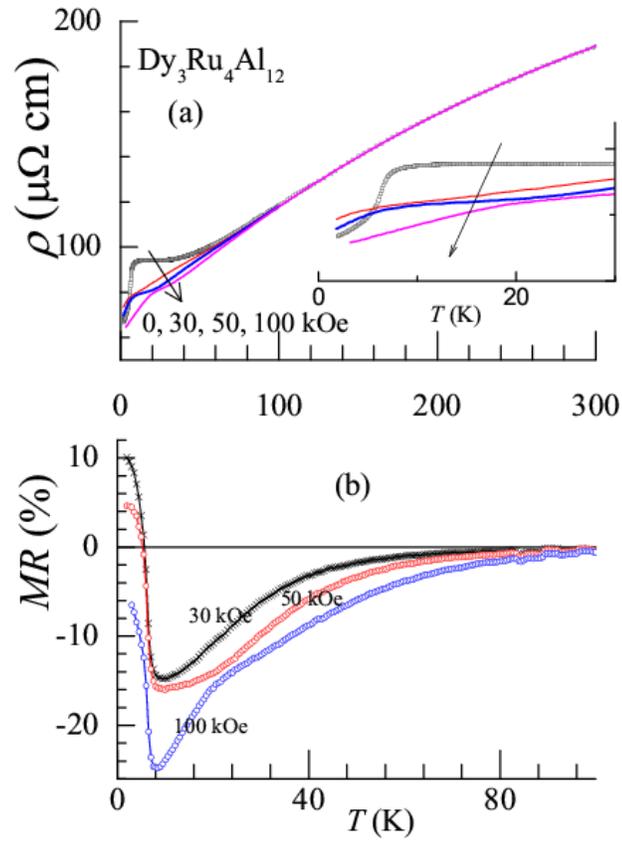

FIGURE 8

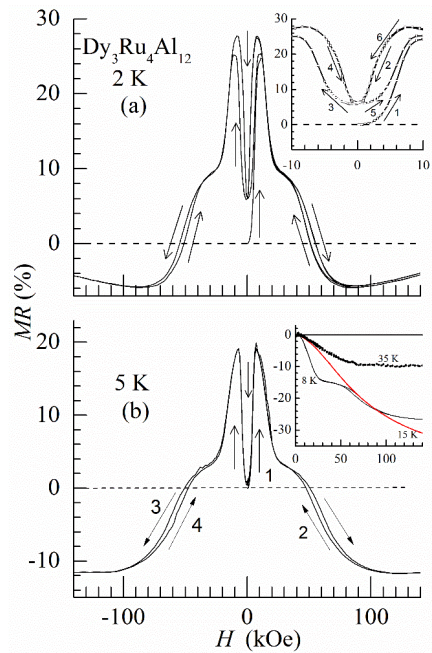

FIGURE 9